\documentclass[pra,twocolumn,showpacs]{revtex4}
\usepackage{graphicx}
\begin{document}

\bibliographystyle{unsrt}

\title{Demonstration of Quantum Error Correction using Linear Optics}
\author{T.B. Pittman, B.C Jacobs, and J.D. Franson}
\affiliation{Johns Hopkins University,
Applied Physics Laboratory, Laurel, MD 20723}

\date{\today}

\begin{abstract}
We describe a laboratory demonstration of a quantum error correction procedure that can correct intrinsic measurement errors in linear-optics quantum gates.  The procedure involves a two-qubit encoding and fast 
feed-forward-controlled single-qubit operations. In our demonstration the qubits were represented by the polarization states of two single-photons from a parametric down-conversion source, and the real-time feed-forward control was implemented using an electro-optic device triggered by the output of single-photon detectors.
\end{abstract}


\maketitle

\section{Introduction}
\label{sec:introduction}

Large-scale quantum computing will require quantum error correction (QEC) to protect qubits from the effects of noise and undesired measurements. The basic approach is to encode a qubit in such a way that any errors can be identified and corrected without measuring the value of the qubit itself  \cite{nielsenchuangbook,shor95}.  Various QEC protocols have recently been demonstrated in NMR \cite{cory98,leung99,knill01a} and ion-trap \cite{chiaverini04} approaches to quantum computing.  In linear optics quantum computing (LOQC), the most common error consists of a measurement of the value of a qubit, which can occur during quantum logic operations \cite{knill01b}. Measurement errors of that kind can be corrected using two-qubit encoding combined with fast feed-forward control \cite{knill01b}.  Several encoding \cite{pittman04a,zhao04,obrien04} and feed-forward experiments \cite{giacomini02,pittman02a,brida04,ursin04} have recently been reported. In this paper we combine these two techniques to demonstrate QEC for measurement errors in LOQC.  

In LOQC, failures in the probabilistic logic gates correspond to situations in which the value of a single-photon qubit is measured in the computations basis (a Z-measurement) \cite{knill01b}. The same situation applies to the recent Zeno gate approach as well \cite{franson04}. However, these intrinsic measurement errors can be corrected by using the following two-qubit encoding \cite{knill01b,obrien04}:

\begin{eqnarray}
\nonumber |0\rangle \rightarrow |0_{L}\rangle 
\equiv \frac{1}{\sqrt{2}}(|00\rangle +|11\rangle)\\ 
|1\rangle \rightarrow |1_{L}\rangle
\equiv \frac{1}{\sqrt{2}}(|01\rangle +|10\rangle)
\label{eq:twoqubitcode}
\end{eqnarray}

\noindent In this code a single-photon qubit with value 0 (or 1) is encoded into a logical qubit represented by the two-photon Bell state $\phi^{+}$ (or $\psi^{+}$). The value of the logical qubit corresponds to the parity of the two physical qubits.  The same encoding must also be applied to superposition states. 

From equations (\ref{eq:twoqubitcode}) it can be seen that if a Z-measurement occurs on either of the two photons, and the value 0 is found, the state of the remaining photon simply corresponds to that of the initial single-photon qubit.  On the other hand, if the Z-measurement results in the value 1, the state of the remaining photon corresponds to the bit-flipped value of the initial qubit.  In this latter case, a fast feed-forward-controlled bit-flip is used to restore the original qubit value. 

\begin{figure}[b]
\includegraphics[angle=-90,width=3in]{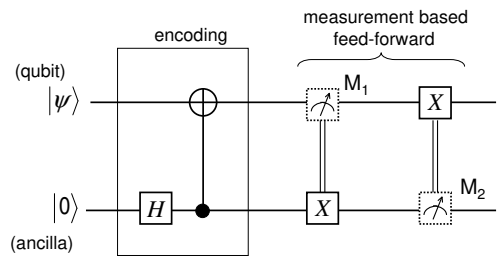}
\vspace*{-.75in}
\caption{Quantum circuit diagram \protect\cite{nielsenchuangbook} illustrating protection and recovery from a Z-measurement error in LOQC \protect\cite{knill01b}. An ancilla photon and encoding operation are used to convert a single-photon qubit $|\psi\rangle$ into the two-photon logical qubit $|\psi_{L}\rangle$ according to equation (\protect\ref{eq:twoqubitcode}). The dashed-boxes $M_{1}$ and $M_{2}$ symbolize Z-measurements which may or may not occur.  If one does occur, and returns the value 1, a bit-flip ($X$) is applied using feed-forward control.  This procedure recovers the initial qubit $|\psi\rangle$. If needed, the logical qubit $|\psi_{L}\rangle$ can then be regenerated by supplying a new ancilla photon and repeating the encoding operation.}
\label{fig:overview}
\end{figure}

A quantum circuit diagram \cite{nielsenchuangbook} illustrating this encoding and feed-forward-control is shown in Figure \ref{fig:overview}.  A single-photon qubit in an arbitrary state $|\psi\rangle$ and an ancilla photon in the state $|0\rangle$ are sent into an encoding device which produces the two-photon logical qubit $|\psi_{L}\rangle$.  If an unwanted Z-measurement $M_{1}$ occurs on bit 1, and the value 1 (or 0) is found, the state of the photon in the lower path is bit-flipped (or left alone) to recover the original qubit $|\psi\rangle$. The same procedure is used to recover $|\psi\rangle$ in the event of a measurement $M_{2}$ on bit 2 \cite{m1andm2}.  In either case, the two-photon logical qubit $|\psi_{L}\rangle$ can be recovered by regenerating a new ancilla photon and repeating the encoding process.

In our demonstration of the QEC technique of Figure \ref{fig:overview}, the qubit $|\psi\rangle$ and ancilla state $|0\rangle$ were represented by the polarization states of two single-photons from a parametric down-conversion pair.  The encoding to produce $|\psi_{L}\rangle$ was done probabilistically using linear optics and post-selection, and the 
feed-forward-controlled bit-flip was accomplished using an electro-optic polarization rotator (Pockels cell) triggered by the output of single-photon detectors. 

The basic idea of the experiment was to intentionally inflict a Z-measurement on one of the photons, and then verify the success of the QEC procedure by comparing the corrected polarization state of the remaining output photon with the input state $|\psi\rangle$. The subsequent repetition of the encoding operation to regenerate the two-photon logical qubit $|\psi_{L}\rangle$ was not included in this demonstration.

\section{Encoding in the Coincidence Basis}
\label{sec:encoding}

Generating the two-qubit code of equation (\ref{eq:twoqubitcode}) requires a non-trivial entangling operation between the qubit and ancilla photons. In principle, operations of this kind can be performed near-deterministically in LOQC by incorporating large numbers of additional photons and very high-efficiency detectors \cite{knill01b}. 

For laboratory demonstrations, however, these requirements can be greatly reduced by working in the so-called ``coincidence basis'', which utilizes destructive measurements to ensure that photons were actually present in the desired optical paths \cite{ralph02}.  In many cases, this simplification can be used to successfully demonstrate the essential features of a two-qubit logic operation while overcoming the effects of random photon sources, loss, and limited detector efficiency associated with current technology.  For example, a coincidence-basis photonic CNOT gate \cite{obrien03} was recently used to demonstrate the encoding (and decoding) of equation (\ref{eq:twoqubitcode}) \cite{obrien04}.

The encoding box of Figure \ref{fig:overview} can be further simplified by exploiting the fact that the ancilla photon is always in the fixed state $|0\rangle$. This allows one to use linear optics to construct a robust specific-purpose encoding device that does not require the general functionality of a full CNOT gate. In our experiment with polarization qubits, this encoding was done using  a single polarizing beam splitter (PBS). The use of a PBS to implement two-qubit logic operations has been done in other contexts as well (see, for example, \cite{pittman02b,pan03}). 

A PBS is a four-port device that transmits horizontally polarized single photons ($|H\rangle$) and reflects vertically polarized single photons ($|V\rangle$). We use the following polarization definitions for the computational basis: 
$|0\rangle \equiv \frac{1}{\sqrt{2}}(|H\rangle + |V\rangle)$ (a photon polarized at $45^{o}$), and $|1\rangle \equiv \frac{1}{\sqrt{2}}(|H\rangle - |V\rangle)$ (a photon polarized at $-45^{o}$). 

A qubit photon in an arbitrary state $|\psi\rangle = \alpha |0\rangle + \beta |1\rangle$ is sent into one of the input ports of of the PBS, while the ancilla photon (in fixed state $|0\rangle$) is sent into the second input port of the PBS.  Provided that one photon exits each output port, it can be shown that the two-qubit code of equation (\ref{eq:twoqubitcode}) is achieved:

\begin{equation}
\alpha |0\rangle + \beta |1\rangle \rightarrow 
\frac{\alpha}{\sqrt{2}} (|00\rangle +|11\rangle) + \frac{\beta}{\sqrt{2}}  (|01\rangle +|10\rangle)
\label{eq:qubitencoding}
\end{equation}

\noindent For any qubit value $|\psi\rangle$, the probability that one photon will exit each output port is $\frac{1}{2}$. This can be viewed as the ideal success probability of this probabilistic encoding device. In our experiment, coincidence basis measurements were used to monitor only those cases in which that occurred.

\section{Error Correction Experiment}
\label{sec:experiment}

\begin{figure}[b]
\vspace*{.2in}
\includegraphics[angle=-90,width=3.5in]{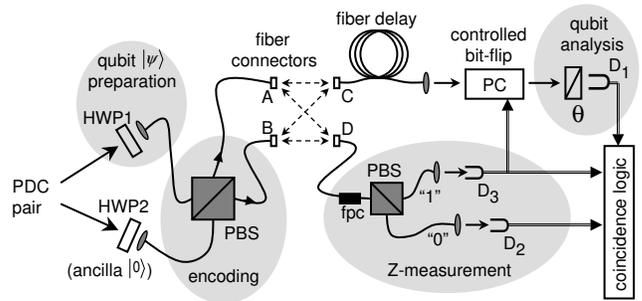}
\vspace*{-1in}
\caption{Apparatus used to demonstrate LOQC QEC.  The shaded areas relate key aspects of the apparatus to the quantum circuit diagram of Figure \protect\ref{fig:overview}.  Details and symbols are described in the text.}
\label{fig:experiment}
\end{figure}

An overview of the QEC experiment is shown in Figure \ref{fig:experiment}. The shaded areas are used to relate several key aspects of the apparatus to the quantum circuit diagram of Figure \ref{fig:overview}: qubit $|\psi\rangle$ preparation, the encoding device described in section \ref{sec:encoding}, a Z-measurement with feed-forward control, and qubit analysis to verify the QEC procedure.

A type-I down-conversion source (PDC) produced pairs of horizontally polarized photons at 780nm that were used as the qubit and ancilla (source details can be found in \cite{pittman04a}).  A half-wave plate (HWP2) was used to fix the polarization state of the ancilla photon at $45^{o}$ (logical $|0\rangle$), while a rotatable half-wave plate (HWP1) could be used to prepare different linear polarization qubit states $|\psi\rangle$.

The qubit and ancilla photons were injected into a single-mode fiber-coupled PBS for the encoding. For a general qubit value $|\psi\rangle$, the encoding operation can essentially be understood as a two-photon quantum interference effect that uses a beam splitter and post-selection to generate polarization entanglement (in the coincidence basis) from an initial product state of two single photons \cite{shih88}. This required the photons to arrive at the PBS within a time defined by their coherence lengths, and alignment of the encoder involved optimizing various polarization-dependent Hong-Ou-Mandel-type quantum interference effects  \cite{hong87}. The fidelity of the encoded logical qubit $|\psi_{L}\rangle$ was directly related to the quality of these two-photon interference effects. This, in turn, impacted the ability to recover the initial single qubit state $|\psi\rangle$ after a Z-measurement on one of these photons.

Fibers A and B containing the two-photon logical qubit $|\psi_{L}\rangle$ were connected to fibers D and C which led, respectively, to the Z-measurement device and to the feed-forward-controlled bit-flip and output qubit analysis zone. As shown by the dashed arrows in the figure, these fiber connections could be easily swapped to make the connections A:C and B:D, or A:D and B:C.  This allowed us to make a Z-measurement on either of the photons comprising the logical qubit $|\psi_{L}\rangle$, and then correct the state of the remaining photon to recover $|\psi\rangle$. 

The Z-measurement was accomplished using a second fiber-coupled PBS.  A fiber polarization controller (fpc) was used to rotate the alignment of the transmission/reflection axes of this PBS into the computational basis. In this way, a photon with polarization corresponding to the state $|1\rangle$ would be transmitted by the PBS to a single-photon detector $D_{3}$, while a photon with polarization corresponding to the state $|0\rangle$ would be reflected by the PBS to a single-photon detector $D_{2}$.  All detectors were preceded by 10nm bandpass interference filters centered at 780nm.

The feed-forward-controlled bit-flip was implemented using a Pockels cell (PC) that was triggered only by the output of detector $D_{3}$.  Additional technical details about this part of the experiment can be found in our earlier work on feed-forward control \cite{pittman02a}. Here the PC was oriented with its fast axis in the horizontal direction, so that the application of a half-wave voltage pulse triggered by $D_{3}$ would cause a bit-flip in the computational basis.   

Because this feed-forward control process took roughly 100 ns \cite{pittman02a},  a 30 m fiber delay line was used to delay the output photon before entering the PC. The polarization state of the corrected (or uncorrected) photons exiting the PC were then measured using a rotatable polarization analyzer $\theta$ and detector $D_{1}$.  A coincidence logic circuit was used to record only those events in which one photon was detected by the Z-measurement detectors, and the second photon was detected by $D_{1}$. This enforced the required coincidence-basis operation of the encoding device by rejecting those cases in which both photons of a PDC pair exited the same port of the encoding PBS.

\section{Results}
\label{sec:results}

\begin{figure*}[t]
\vspace*{-.25in}
\includegraphics[angle=-90,width=6in]{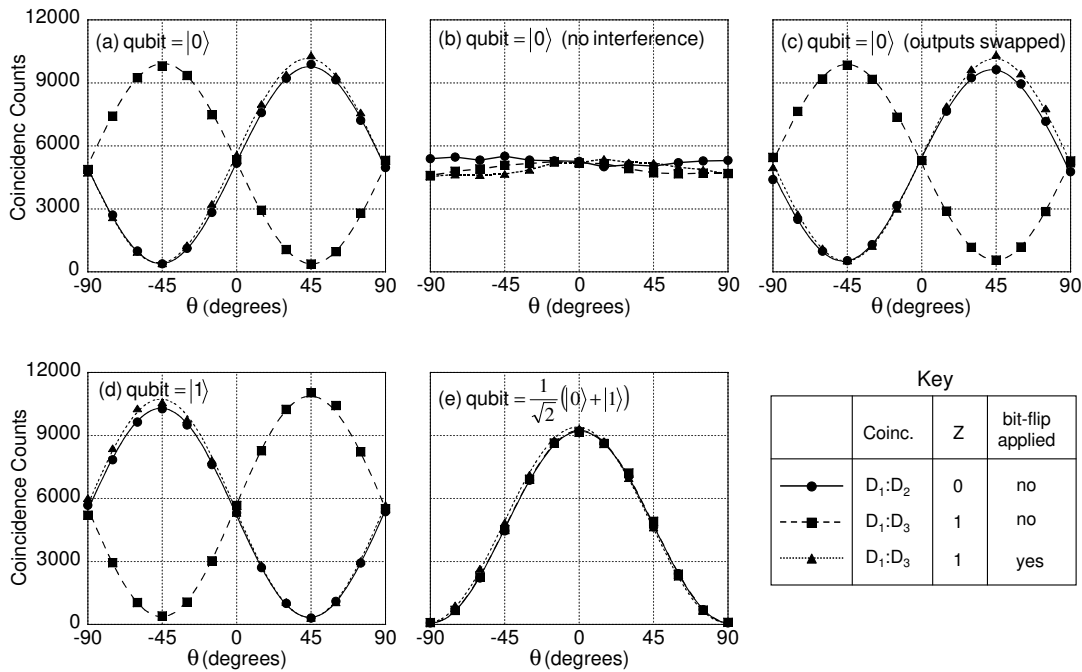}
\vspace*{-.75in}
\caption{Experimental results of the QEC procedure for several different values of input qubits $|\psi\rangle$. The data show the number of coincidence counts per 60 seconds as a function of the output qubit polarization analyzer $\theta$. The Key summarizes the experimental conditions for the three sets of data shown in each plot. In (a),(c),(d), and (e) the lines are sinusoidal fits to the data with an average visibility of 93.5\%. The corrected qubit examples showed the ability to fully recover the qubit $|\psi\rangle$ from a Z-measurement error with an estimated average fidelity of about 98\%.}
\label{fig:data} 
\end{figure*}

In practice, the demonstration of quantum error correction consisted of using HWP1 to specify a qubit value $|\psi\rangle$, and then monitoring the coincidence counting rate between detectors $D_{1}$ and $D_{2}$, or $D_{1}$ and $D_{3}$, as a function of the analyzer angle $\theta$.  The results obtained for several different examples of $|\psi\rangle$ are shown in Figure \ref{fig:data}.

For the data shown in Figure \ref{fig:data}(a), the HWP1 was used to prepare the qubit in the state $|0\rangle$ (a photon polarized at $45^{o}$), and the fiber connections A:C and B:D were used. 

The solid-circle data points show the number of $D_{1}:D_{2}$ coincidence counts  as a function of $\theta$. This corresponds to a Z-measurement returning the value 0. In accordance with equation (\ref{eq:twoqubitcode}), no correction is needed, and the output qubit is expected to be in the same state as the input $|\psi\rangle$. The results agree with this prediction; a Malus' law dependence on $\theta$ consistent with a linear polarization state of $45^{o}$ is clearly seen.  The solid line is a sinusoidal fit to the data points, with a visibility of (92.2 $\pm$ 0.3)\%. The data points obtained with the polarizer $\theta$ set to $\pm 45^{o}$ (corresponding to the qubits $|0\rangle$ and $|1\rangle$) can be used to estimate a recovery of $|\psi\rangle = |0\rangle$  from the Z-measurement error with a fidelity of  $\sim 98\%$.

The solid-square data points in Figure \ref{fig:data}(a) show the number of $D_{1}:D_{3}$ coincidence counts as a function of $\theta$. This corresponds to a Z-measurement resulting in the value 1, which requires a feed-forward controlled bit-flip on the output qubit to recover $|\psi\rangle = |0\rangle$. Here, however, the PC was intentionally disconnected so that no bit-flip was applied.  As expected, the data is indicative of an output qubit $|1\rangle$. In this example, the dashed line fit to the data had a visibility of (92.5 $\pm$ 0.3)\%.

The solid-triangle data in Figure \ref{fig:data}(a) correspond to the same situation, but with the PC connected and the bit-flip applied.  In this case the data clearly shows the recovered qubit $|\psi\rangle = |0\rangle$. The dotted-line fit to this data has a visibility of (92.1 $\pm$ 0.3)\%, indicating the ability to successfully apply the feed-forward-controlled bit-flip using the PC.

From an experimental point of view, a comparison of the results in Figures \ref{fig:data}(a) and \ref{fig:data}(b) highlights the non-classical nature of the encoding operation.  The conditions were exactly the same for these two data sets, except that for (b) the qubit photon was delayed relative to the ancilla photon by roughly twice its coherence length before entering the encoding PBS. This temporal information rendered the two photons distinguishable, which destroyed the quantum interference effects necessary for successful encoding \cite{hong87}. The resulting flat lines in Figure \ref{fig:data}(b) are what would be expected from ``classical'' statistics in this case : roughly half of the photons emerging in fiber A were horizontally polarized, and the other half were vertically polarized.

The data shown in Figure \ref{fig:data}(c) corresponds to the conditions of Figure \ref{fig:data}(a), except that the outputs of the encoding device were swapped by making the fiber connections A:D and B:C.  The three sets of data are nearly identical to those in Figure \ref{fig:data}(a), which demonstrates the ability of the two-qubit code (\ref{eq:twoqubitcode}) to recover from a Z-measurement on either of the two photons comprising the logical qubit $|\psi_{L}\rangle$.

For the data shown in Figures \ref{fig:data}(d) and \ref{fig:data}(e) the HWP1 was used to prepare input qubits $|\psi\rangle$ in the states $|1\rangle$ and $\frac{1}{\sqrt{2}}(|0\rangle + |1\rangle$). In both cases, the Malus' law dependence on $\theta$ agrees with the expected output states.  Similar results were also obtained with the output fibers swapped. The average visibility of the three fits in Figure \ref{fig:data}(d) was 93.6\%, while it was 98.2\% for Figure \ref{fig:data}(e). The higher visibility in the latter case is due to the fact that the encoding operation does not depend on two-photon interference effects for the superposition state $|\psi\rangle = \frac{1}{\sqrt{2}}(|0\rangle + |1\rangle)$, which simply corresponds to a horizontally polarized photon. The slight deviation from 100\% visibility in this case can therefore be used to estimate the magnitude of the remaining technical errors in the experiment.

The average visibility of the nonclassical two-photon interference patterns corresponding to the corrected qubits in Figures \ref{fig:data}(a),(c),(d), and (e) was 93.6\%; in contrast, the visibilities in Figure \ref{fig:data}(b) (with no quantum interference) were essentially zero. These examples clearly show the ability to recover the qubit $|\psi\rangle$ from a Z-measurement error.

\section{Discussion}
\label{sec:Discussion}

The results of Figure \ref{fig:data} demonstrate the ability to combine two-qubit encoding and feed-forward control to recover from the Z-measurement errors intrinsic in probabilistic LOQC logic gates.  

Within the coincidence basis, the two-qubit encoding of equation (\ref{eq:twoqubitcode}) can be accomplished using a photonic CNOT gate, as was first demonstrated in reference \cite{obrien04}, or by using a specific purpose encoding device such as the one constructed here.  The required feed-forward control was implemented here using real-time polarization rotations via the techniques of reference \cite{pittman02a}.

All of these preliminary studies have shown that intrinsic error correction is feasible in an LOQC approach. However, it should be noted that the two-qubit code and feed-forward used here cannot correct for more general physical errors such as bit-flips, phase-shifts, and loss.  In order to overcome errors of that kind, the procedure demonstrated here would need to be embedded in a more general QEC code \cite{knill01b}. The realization of these more complex codes, as well as operation outside the coincidence basis, will place significant demands on the photon sources and detectors.

This work was supported by ARO, ARDA, and IR\&D funding.




\begin{thebibliography}{50}

\bibitem{nielsenchuangbook} {\em Quantum Computing and Quantum Information}, M.A.
Nielsen and I.L. Chuang, Cambridge University Press (2000).

\bibitem{shor95}  P.W. Shor, Phys. Rev. A {\bf 52}, 2493 (1995).

\bibitem{cory98} D.G. Cory {\em et.al}, Phys. Rev. Lett. {\bf 81}, 2152 (1998).

\bibitem{leung99} D. Leung {\em et.al}, Phys. Rev. A {\bf 60}, 1924 (1999).

\bibitem{knill01a} E. Knill, R. Laflamme, R. Martinez, and C. Negrevergne, Phys. Rev. Lett. {\bf 86}, 5811 (2001).

\bibitem{chiaverini04} J. Chiaverini {\em et.al.}, Nature {\bf 432}, 602 (2004).

\bibitem{knill01b} E. Knill, R. Laflamme, and G.J. Milburn, Nature {\bf 409}, 46 (2001).

\bibitem{pittman04a} T.B. Pittman, B.C. Jacobs, and J.D. Franson, Phys. Rev. A {\bf 69}, 042306 (2004). 

\bibitem{zhao04} Z. Zhao {\em et.al}, Nature {\bf 430}, 54 (2004).  

\bibitem{obrien04} J.L. O'Brien, G.J. Pryde, A.G. White, and T.C. Ralph, quant-ph/0408064 (2004).

\bibitem{giacomini02} S. Giacomini, F. Sciarrino, E. Lombardi, and F. De Martini, Phys. Rev. A {\bf 66}, 030302 (2002)

\bibitem{pittman02a} T.B. Pittman, B.C. Jacobs, and J.D. Franson, Phys. Rev. A {\bf 66}, 052305 (2002). 

\bibitem{brida04} G. Brida {\em et.al} Phys. Rev. A {\bf 70}, 032332 (2004). 

\bibitem{ursin04} R. Ursin, {\em et.al} Nature {\bf 430}, 849 (2004). 

\bibitem{franson04} J.D. Franson, B.C. Jacobs, and T.B. Pittman, Phys. Rev. A {\bf 70}, 062302 (2004).

\bibitem{m1andm2} This proceedure cannot correct for errors in the less-likely event of simultaneous measurments $M_{1}$ and $M_{2}$.


\bibitem{ralph02} T.C. Ralph, N.K. Langford, T.B. Bell, and A.G. White, 
Phys. Rev. A {\bf 65}, 062324 (2002). 

\bibitem {obrien03} J.L. O'Brien, G.J. Pryde, A.G. White, T.C. Ralph, and D. Branning, Nature {\bf 426}, 264 (2003).  

\bibitem{pittman02b} T.B. Pittman, B.C. Jacobs and J.D. Franson, Phys. Rev. Lett. {\bf 88}, 257902 (2002).  

\bibitem{pan03}J.-W. Pan, S. Gasparoni, R. Ursin, G.Weihs, and A. Zeilinger 
Nature {\bf 423}, 417 (2003).

\bibitem{shih88} Y.H. Shih and C.O. Alley, Phys. Rev. Lett. {\bf 61}, 2921 (1988).

\bibitem{hong87} C.K. Hong, Z.Y. Ou, and L. Mandel, Phys. Rev. Lett. {\bf 59}, 2044 (1987).








\end{thebibliography}
\end{document}